\documentclass[preprint]{aastex}

\begin{document}

\title{The Light Curve of the Weakly-Accreting T Tauri Binary KH 15D from 2005-10: Insights into the Nature of its Protoplanetary Disk}

\author{ William Herbst$^1$, Katherine LeDuc$^1$, Catrina M. Hamilton$^2$, Joshua N. Winn$^3$, Mansur Ibrahimov$^4$, Reinhard Mundt$^5$ and Christopher M. Johns-Krull$^6$}

\affil{$^1$Astronomy Department, Wesleyan University, Middletown, CT 06459 USA}
\affil{$^2$Physics and Astronomy Department, Dickinson College, Carlisle, PA 17013 USA}
\affil{$^3$Department of Physics, and Kavli Institute for Astrophysics
and Space Research, Massachusetts Institute of Technology,
Cambridge, MA 02139}
\affil{$^4$Ulugh Bek Astronomical Institute of the Uzbek Academy of Sciences, Astronomicheskaya 33, Tashkent 700052 Uzbekistan}
\affil{$^5$Max-Planck-Institut f\"ur Astronomie, K\"onigstuhl 17, D-69117, Heidelberg, Germany}
\affil{$^6$Department of Physics and Astronomy, Rice University, Houston, TX 77005 USA}
\email{wherbst@wesleyan.edu}

\begin{abstract}

Photometry of the unique pre-main sequence binary system KH 15D is presented, spanning the years 2005-2010. This system has exhibited photometric variations and eclipses over
the last $\sim$50 years that are attributed to the effect of a precessing
circumbinary disk. Advancement of the occulting edge across the projection on the sky of the binary orbit has continued and the photospheres of both stars are now completely obscured at all times. The system has thus transitioned to a state in which it should be visible only by scattered light, and yet it continues to show a periodic variation on the orbital cycle with an amplitude exceeding two magnitudes. This variation, which depends only on the binary phase and not on the height of either star above or below the occulting edge, has likely been present in the data since at least 1995. It can, by itself, account for the ``shoulders" on the light curve prior to ingress and following egress, obviating to some degree the need for components of extant models such as a scattering halo around star A or forward scattering from a fuzzy disk edge. However, the spectroscopic evidence for some direct or forward scattered light from star A even when it was several stellar radii below the occulting edge shows that these components can probably not be fully removed, and raises the possibility that the occulting edge is currently more opaque than it was a decade ago, when the spectra were obtained. 
A plausible source for the variable scattering component is reflected light from the far side of a warped occulting disk. We have detected color changes in V-I of several tenths of a magnitude to both the blue and red that occur during times of minima. These may indicate the presence of a third source of light (faint star) within the system, or a change in the reflectance properties of the disk as the portion being illuminated varies with the orbital motion of the stars. The data support a picture of the circumbinary disk as a geometrically thin, optically thick layer of perhaps mm or cm-sized particles that has been sculpted by the binary stars and possibly other components into a decidedly nonplanar configuration. A simple (infinitely sharp) knife-edge model does a good job of accounting for all of the recent (2005-2010) occultation data when one allows for the scattered light component, variations from cycle to cycle in the location of the edge at the level of 0.1-0.2 stellar diameters and the possible alignment of the spot distribution on star A with its projected orbital motion.   
\end{abstract}
 
\keywords{stars: pre-main sequence - stars: individual - KH 15D}
 
\section{Introduction}

KH 15D is a binary system composed of $\sim$0.5 M$_\odot$ pre-main sequence stars in a highly eccentric ($e \approx 0.6$) orbit with a period of 48.37 days \citep{j04,Ham05,w04,w06}. The orbital plane is inclined to an opaque circumbinary disk or ring that precesses on a time scale of $\sim$1000 years  \citep{CM04} and has been progressively occulting the orbits of the stars since about 1960 \citep{J04}. This has produced dramatic photometric variability that has evolved steadily and been vigilantly monitored since its re-discovery in the mid-1990's \citep{KH98,Her02,Ham05}. Here we report on the last five years of this evolution during which time the disk edge has proceeded to completely cover the orbit and photosphere of star A. Star B was last seen in 1995, so the system has now entered a state of diminished optical brightness in which light from the illuminating binary reaches us only though scattering from circumstellar or circumbinary matter. It is uncertain how long this state will persist, perhaps for centuries or perhaps for only a brief time.

The last few years of the progression of the occulting edge have provided us with information that helps constrain properties of the disk, the binary orbit, star A and its immediate environment. In particular, we can study the vertical distribution of the obscuring matter that composes the disk, its optical properties, and the details of its progression across the orbit. Because the edge is so sharp -- a small fraction of a stellar radius (see below) -- we can also seek constraints on the (inhomogeneous) surface flux and temperature distribution across star A and on the elements of the binary orbit. This information is largely of a sort that cannot be obtained for any other star or binary system, pre-main sequence or otherwise. Hence, KH 15D plays a unique and important role in studies of pre-main sequence stars, as well as disk evolution. 

We emphasize that there is no reason to expect that the stars or disk in KH 15D are in any way unusual for T Tauri stars. What is unusual is the current alignment of the disk with respect to our line of sight. It is, therefore, reasonable to suppose that what we learn about the physical state of this system, in particular its disk, by exploiting the fortuitous geometry will reliably inform studies of protoplanetary disks in general. With a total mass near 1 M$_\odot$ and an age of $\sim$3 My, KH 15D may be particularly relevant to the stage of planet formation in our own Solar System during which the most primitive meteorites and their embedded chondrules formed. \citet{p08,p10} have recently identified two other systems with KH 15D-like variations and have proposed that they form a new class of variable stars with warped inner disks caused by the existence of tertiary components. There is possible evidence for third light in the photometry of KH 15D presented in this paper, in support of that idea. 

This paper presents photometric data in V and I over the period 2005-10 and an empirical description and basic analysis of those data. It will hopefully be useful to a next generation of dynamical and phenomenological models along the lines of those proposed by \citet{CM04}, \citet{w04,w06} and \citet{sa08}. It is clear that none of the current models accurately predicted the observed behavior reported here, nor do any of them incorporate all of the relevant physical components, such as a scattered light source with the observed properties. It is likely that the next generation of models will make it possible to exploit more fully and confidently the information in the data presented here than we can do empirically. The current paper will thus be restricted to presenting the data and providing some qualitative guidance to its interpretation based on the perspective of observers.

\section{Observations}

In this paper we report V and I (on the Cousins system) photometric data on KH 15D obtained at the Van Vleck Observatory (VVO) on the campus of Wesleyan University in Middletown, CT, at Cerro Tololo Inter-American Observatory (CTIO) in Chile with a SMARTS consortium telescope and from Mount Maidanak Observatory (MMO) in Uzbekistan, operated by the Uzbek Academy of Sciences. Five years of observations from September, 2005, through March, 2010, are presented. 

At VVO, data were obtained with the 0.6 m Perkin telescope. In 2005-6 and 2006-7 we employed a 1024 x 1024 front-illuminated Photometrics CCD camera. Five successive one-minute exposures in I were taken once each night and summed. Since 2007-8 the CCD camera has been a 2048 x 2048 back-illuminated Apogee camera with 0.3$\arcsec$ pixels. Between five and fifteen consecutive 1-minute duration I or V images were obtained on clear nights with this system and summed. Typical seeing at VVO is about 2.5$\arcsec$ and the sky is rather bright, so the data are more precise by a factor of 2-3 during the brighter portions of the system's range than during the fainter .

At CTIO, images were obtained each clear night in all five observing seasons from the beginning of October until the end of March with a 2048 x 2048 Fairchild CCD attached to the 1.3 m telescope of the SMARTS consortium. All images employed 2 x 2 binning and had a 6.3$\arcmin$ x  6.3$\arcmin$ field of view. Four images of 5 s each were taken while KH 15D was in the bright phase and normally eight images of 10 s duration were taken while it was faint, during the time period October 2005 through Feb. 2006. After that, the duration of the exposures when the star was faint was increased to 150 s in both V and I and normally 4 consecutive images were obtained and summed.

Observations at MMO were obtained in the I band using the 0.6 m and 1.5 m telescopes and a 512 x 512 and 4096 x 4096 CCD camera, respectively. A range of 5 to 20 images was taken each clear night depending on the phase of the system. Exposures were normally 300 s in length. 

Photometry of the VVO and CTIO images was done at Wesleyan, while photometry of the MMO images was done at the Uzbek Academy of Sciences. In each case, standard IRAF tasks for aperture photometry were employed. KH 15D and several comparison stars were measured. The precise sizes of the aperture and annulus chosen for the photometry varied according to the instrument employed. Details are given in Table~\ref{PhotParm}. Comparison stars were chosen from the set labeled A-G by \citet{Ham05} as well as 16D, 17D and 31D from \citet{KH98}. In each season, potential comparison stars were evaluated for variability and the most stable objects chosen. The principal comparison star for most of the data reported here was star F of \citet{Ham05} and its constancy was checked by reference to star C; variations were $\sim$0.01 mag or less. Table~\ref{CompMags} gives the adopted V and I magnitudes for the comparison objects based on the photometry of \citet{Ham05}. The accuracy of the photometry varies substantially depending on the observatory, observing conditions, exposure length, etc. Generally speaking, however, the data are accurate to 0.02 mag or less when the star is at or near its bright phase (I = 14.5 mag) and accurate to 0.05 mag or better when faint. 

Table~\ref{Data} gives the photometric results from the five observing seasons 2005-2010 reported here combined with previously published CCD data from the period 1995-2005 \citep{Ham05}. The full table is available in the electronic version of the journal article. This represents the full set of CCD data obtained by our group on the Cousins system. It does not include SMARTS data transformed to that system that were discussed and analyzed by \citet{Ham05}, nor does it include the photographic photometry that has been reported by \citet{J04}, \citet{J05}, and \citet{M05}. 

In addition to the photometry, we include in the table the parameter $\Delta$X from Model 3 of \citet{w06} through the 2008-9 season. This is a prediction, based on the orbital and other parameters of that model, of the elevation of the center of star A above (-) or below (+) the edge of the occulting disk in units of the radius of star A (R$_A$ = 1.3 R$_\odot = 9 \times 10^8$ m) from \citet{w06}. We show in what follows that this model is no longer an adequate representation of the data, and we caution that the quantity $\Delta$X is progressively less successful in accounting for the photometry as one moves from 2005 through 2009. It is included in Table~\ref{Data} because we needed it to assess the validity of the extant models and in other aspects of the analysis. 

\section{Results}

\subsection{Overview of the Light Curve and its Features}

\begin{figure}
\epsscale{1.0}
\includegraphics{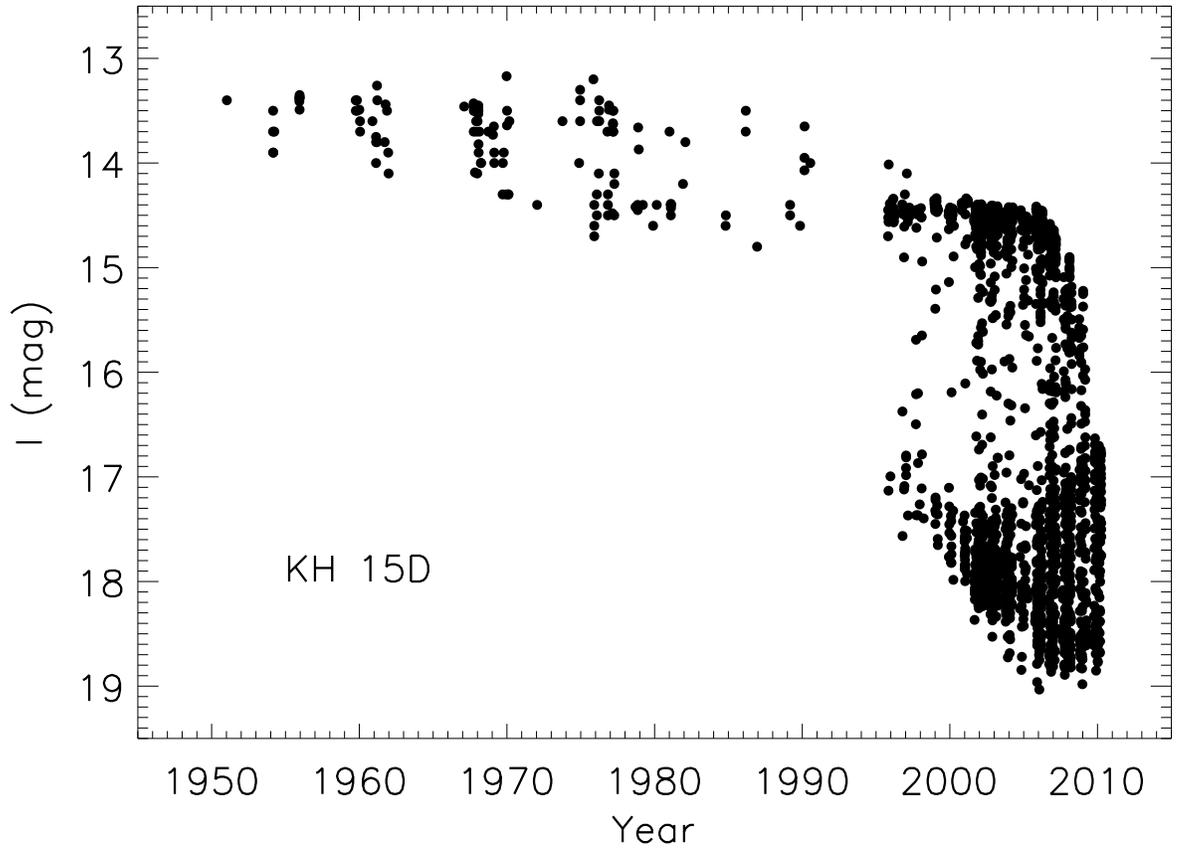}
\caption{The light curve of KH~15D in the Cousins I band. Included here are all of the I or effective I  measurements from the photographic photometry of \citet{J04} and \citet{J05} that span the period 1950 to early 1996 as well as the CCD photometry from \citet{Ham05} and this paper, which cover the period from late 1995 to early 2010 and are expanded in Fig.~\ref{fig1}.}
\label{figh}
\end{figure}

\begin{figure}
\epsscale{1.0}
\includegraphics{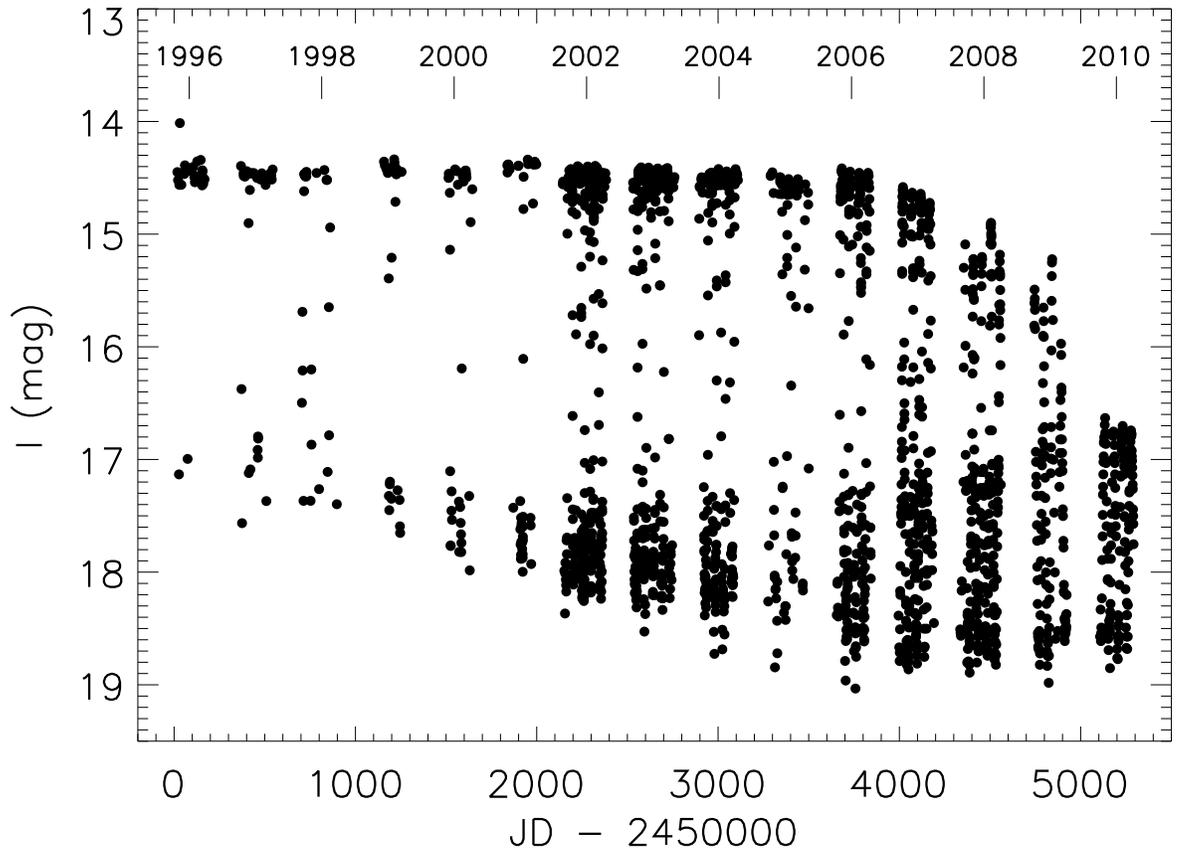}
\caption{The light curve of KH~15D in the Cousins I band based on CCD photometry obtained at VVO, CTIO and MMO from 1995 to 2010. It is quite apparent that the system no longer reaches its previous brightness levels. Note that the one measurement in 1995 showing the system at about magnitude 14 is real. It represents our only fully unobscured CCD detection of Star B.}
\label{fig1}
\end{figure}

\begin{figure}
\epsscale{1.0}
\plotone{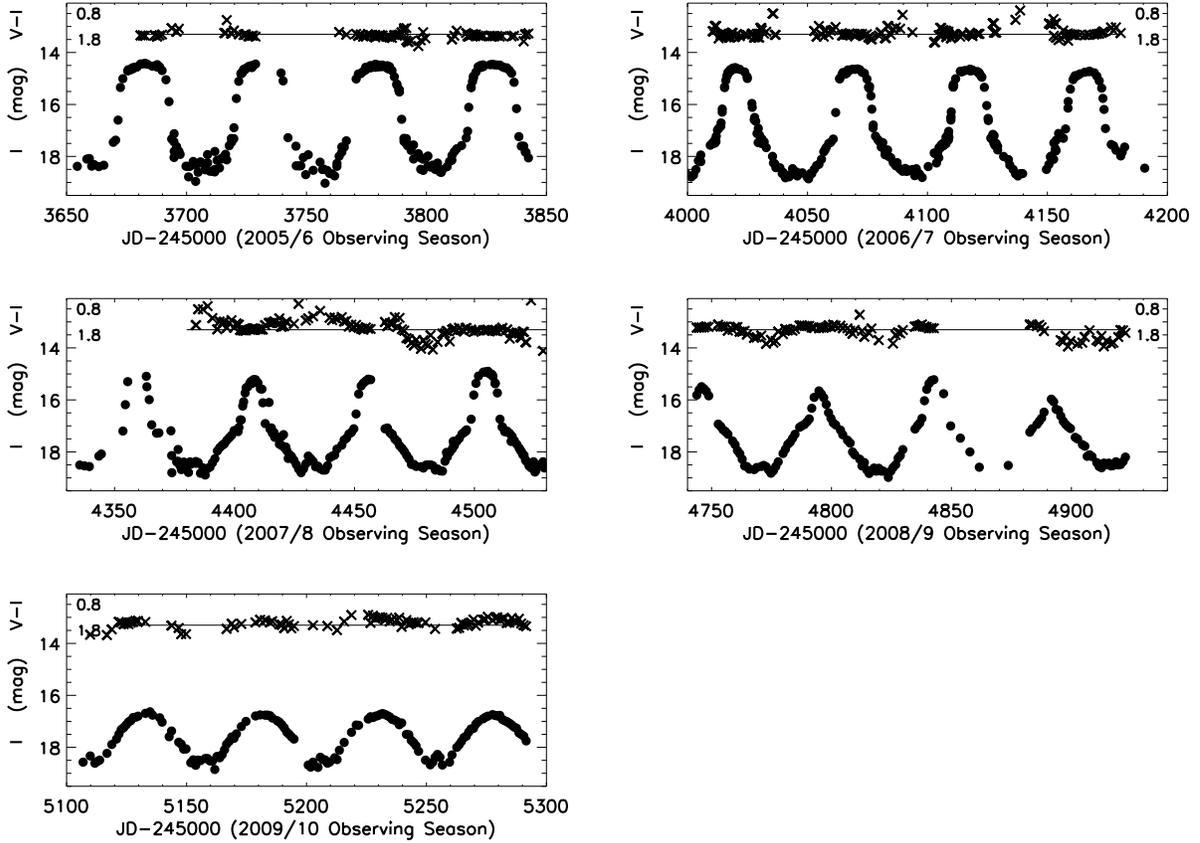}
\caption{The measured I magnitude (solid circles) and V-I color (X's) of KH 15D over the five seasons 2005-6 to 2009-10. The solid line represents the out-of-eclipse mean color of V-I=1.575. Bluer colors are towards the top of the figure. Only color data with errors less than 0.25 mag are plotted. Obviously the time spent near maximum brightness, when star A is not occulted at all, has continued to diminish such that by the 2009-10 observing season star A is fully occulted at all times. It is interesting that color changes are often seen near minimum light but they can be to the blue or to the red and are quite substantial.}
\label{fig}
\end{figure}

\begin{figure}
\epsscale{1.0}
\plotone{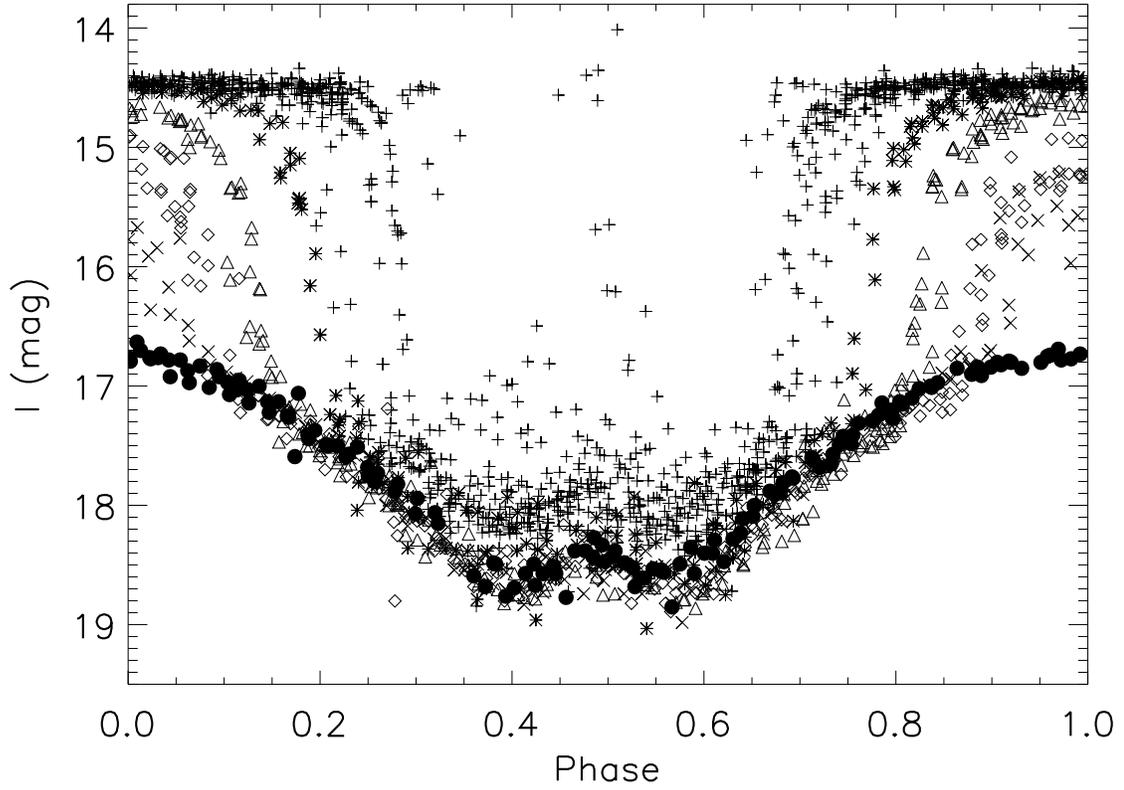}
\caption{The KH~15D light curve phased with the period 48.37 days. Plus signs designate data obtained between 1995 and April of 2005. The new data reported here are plotted as asterisks (2005-6), triangles (2006-7), diamonds (2007-8) and X's (2008-9) and solid circles (2009-10). The light curves show a steady progress towards deeper and broader minima. The 2009-10 data generally define the faint boundary of the light curve and represent the scattered light component of the system alone.}
\label{fig3}
\end{figure}

\begin{figure}
\epsscale{1.0}
\plotone{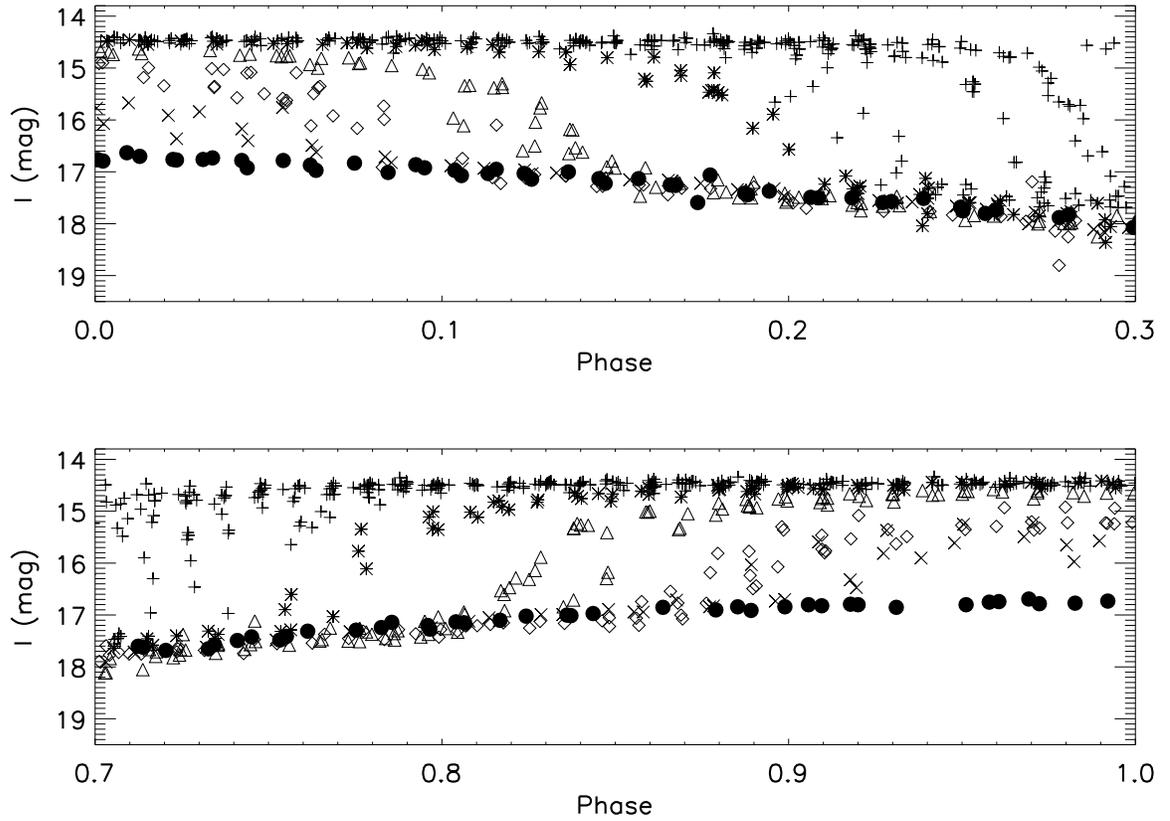}
\caption{Same as Fig. \ref{fig3} but with expanded horizontal axis. It is clearly seen that the light curve in 2009-10 (solid circles), when Star A was fully eclipsed at all binary phases, forms a low luminosity boundary to the light curve at all epochs.}
\label{fig2a}
\end{figure}

Several views of the light curve of KH 15D are provided in Figures~\ref{figh} to \ref{fig2a}. The I magnitude is used because most epochs included I-band observations where the system is relatively bright. Figure~\ref{figh} provides the context for this study by showing all of the I-band photometry available for the star since 1950, which includes the photographic photometry (some of it transformed from optical measurements made at shorter wavelengths) from \citet{J04} and \citet{J05} and the CCD photometry. It is clear that, as predicted, the object has entered into a new phase, one in which neither star is directly visible at any time. We may expect, therefore, that the system brightness will remain below about I = 16.7 for awhile, perhaps even centuries. On the other hand, it is entirely possible that the object will brighten again much sooner than that with the reappearance of either star A or star B. Nothing in the current models allows us to predict how long the total occultation of the binary orbit will persist, other than to say that after a millennium or so we should be back to the current state.

Figure~\ref{fig1} shows all of the CCD photometric data obtained by our group since late 1995 versus time. The one bright point in 1995 represents the only detection in the modern CCD era of star B and indicates that the star is more luminous than star A.\footnote{We adopt the terminology of \citet{w06} to describe the stars: A is the one that has been appearing above the disk edge recently and is, on the basis of its luminosity, of slightly less mass than star B. It may be seen on Fig.~\ref{figh} that there was another similarly bright data point obtained on a different night during the 1995-96 season in the archival photographic record. However, the orbital phase of this datum (a U-band measurement transformed to I-band) is not consistent with its being an actual detection of star B and the error bars on it are too large to be confident that it is even significantly out of line with the rest of the photographic data.} From 1996 to 2006 the system had a peak brightness of around 14.5 mag \citep{Ham05}.

Since 2006 there has been a steady decline in the system's brightness at maximum (see Fig.~\ref{fig1}). In Figure~\ref{fig} we expand the time axis to illustrate the detailed behavior of the star during the last five observing seasons. According to the models of \citet{CM04}, \citet{w04,w06} and \citet{sa08}, the decrease in maximum system brightness is due to the steady advance of the occulting edge across the projection of the binary orbit on the sky. While a decline such as this was predicted in all of the above models, its timing and the extent of the decline was not correctly predicted in any of them. Substantial fading has occurred many years earlier than predicted by \citet{CM04} and about a year later than predicted by \citet{w06}\footnote{This statement refers to Model 3 of \citet{w06} which employs the astrophysical constraint that star A be less massive than star B because it is less luminous. If that constraint is not enforced then one has the Model 1 prediction from that paper, which would have star A partly visible until late 2010, again in disagreement with the observations, but in the opposite sense.}  or \citet{sa08}. Also, the brightness level to which the system has faded upon full occultation of both stars is much brighter at some orbital phases than has been modeled. This indicates the need for a revision to the models involving either the precise properties of the binary orbit or the details of the motion of the opaque edge, or both, as well as reconsideration of the source of the scattered light.

Another point that is apparent on Figure~\ref{fig} and is not accounted for in the current picture of the system is the variation that occurs in both the shape and the brightness of successive maxima as time goes on. This is seen clearly in the 2008-9 observing season (middle right panel of Figure~\ref{fig}). The second maximum observed in that season is actually less bright and of shorter duration than the third maximum. This indicates that the progress of the occulting edge across the binary orbit is not completely steady but that deviations in the location of the edge from one orbital cycle to the next can occur. This was also seen during the 2001-2 season \citep{Her02}. Presumably it is caused by some local fluctuations in the height of the disk edge and we quantify these deviations below in Section 4.3.

An additional new feature evident on Figure~\ref{fig} is the variation that occurs in the shape and amplitude of the light reversal near minimum light. For example, during the second minimum in 2007-8 (middle left panel of Fig. \ref{fig}) we see a rather sharp and bright reversal, as contrasted with the third minimum, which is smoother and shallower. These reversals correlate with the times when star B is closer to the disk edge than star A and may indicate variability of that star or of the geometry of the scattering, or both. There are significant color changes associated with these events that may also provide an important clue to their source (see Section 4.5). 

In Figures~\ref{fig3} and \ref{fig2a} we show another view of the brightness data, namely a phased plot based on the orbital period of 48.37 d. Different symbols are used to represent different epochs of observation. A curious feature that emerges clearly on these plots, and is visible more subtly on Figs \ref{fig1} and \ref{fig}, is that the brightness of the system at the point where star A is just fully eclipsed has changed in progressive fashion with time. In the earlier years, the system had a brightness of about I = 17.5 mag at the point where the rate of fading (during ingress) or brightening (during egress) changed slope (phases $\sim$0.3 and 0.7 on Figs.~\ref{fig3} and \ref{fig2a}). This change of slope marks the point in the light curve where star A is just fully eclipsed. More recently, the change in slope has occurred at brighter and brighter magnitudes, reaching I=16.8 mag in 2008-9 (phases $\sim$0.05 and 0.95 on Fig.~\ref{fig3} and \ref{fig2a}). The most recent data (2009-10), plotted as solid circles on Figs.~\ref{fig3} and \ref{fig2a} trace these inflection points rather faithfully. It is now clear that the brightness of the system at the point that star A is just fully occulted is not a constant, as has previously been assumed, but depends on the binary phase. 

A subtle manifestation of this behavior may be seen on Figure \ref{fig1} where there appears a darkening of the faint parts of the light curve demarcated by a diagonal line running from about I=17.5 mag at JD=2000 to I=16.8 mag at JD = 5000. Its visibility on this plot is due to the fact that the observational cadence for these light curves is rather steady but the decline rate of the system slows abruptly once star A has been fully eclipsed during ingress and speeds up abruptly during egress once the limb of the star appears above the disk edge. We believe that this feature of the light curve is an important clue to the nature of the scattered light and return to its implications and explanation in what follows.

A final feature that is clearly visible on Figures~\ref{figh}, \ref{fig1} and \ref{fig3} is the decline in brightness at minimum light, that was quite dramatic for the first decade of CCD observation and has since leveled off. Its explanation depends on the source of the scattered light and is, therefore, uncertain. We return to the issue after first discussing in more detail the system's interesting color behavior.

\subsection{Color Variations}

Previous work has shown that KH 15D usually became bluer by about 0.1 mag in V-I as it went into eclipse and that this blueness generally persisted throughout the faint state \citep{Ham05,Her08}. While there is a good deal of scatter in the pre-2005 color data when the star is faint, it had been thought that this might reflect more the relative difficulty of doing photometry in V, where the star is very faint and scattered light from a nearby B star very bright, than actual variability of the colors. 

Beginning in March, 2006, we made a special effort to improve the color data by obtaining much longer integrations in V, and the results are displayed in Fig.~\ref{fig}. The V-I color is shown at the top of each panel on the same scale as the brightness and with the horizontal line representing the mean out-of-eclipse color of the system, V-I = 1.575. Bluer colors are towards the top on these plots. It is clear that, when fainter, the system is often slightly bluer than it is during its brighter times but it is now also clear that it sometimes reddens substantially and consistently as one enters the deeper minima. The error in color on the individual data points is around 0.03 mag for I brighter than 17, rising to 0.25 mag for I =19. No data with error bars exceeding 0.25 mag is shown in these plots.

It is interesting that during the first full season of more precise measurements (2006-7; upper right panel of Fig. \ref{fig}) the star never appeared substantially redder than its out-of-eclipse values and seemed to follow the common behavior of previous years by getting much bluer during deep minima. This behavior persisted into the next season (2007-8; middle left panel) for the first two minima that had color measurements and then abruptly changed for the last two minima, showing substantial and significant {\it reddening} as the star faded into a deep minimum. All three of the measured minima in 2008-9 also showed reddening at minimum light (see middle right panel of Fig.~\ref{fig}). We return to a discussion of these interesting color changes in Section 4 of the paper.

\subsection{The Light and Color Variations of KH 15D out of Eclipse}

To investigate the intrinsic variability of star A as well as the possibility of extinction and/or reddening by matter above the disk edge, we consider the photometric results obtained when star A was fully above the disk edge. We use the parameter $\Delta$X calculated from Model 3 of \citet{w06} to determine this. We remind the reader that this model is based on a fit to the pre-2005 data only and its extrapolation to the current epoch results in a systematic error in calculating $\Delta$X that must be accounted for in the analysis to follow. For the pre-2005 data, however, the quantity is accurate and meaningful, although cycle-to-cycle variations in the location of the edge with respect to the star of up to a couple tenths of a stellar radius are possible (see below). Figure~\ref{fig4} displays the results.

It is clear from the top panel of Figure~\ref{fig4} that when star A is more than about 5 stellar radii above the opaque edge it shows no trend whatsoever in brightness with increasing height. However, beginning around $\Delta$X$=-5$ there is a small but significant drop in the system brightness as the star approaches the disk edge. This had been noticed before and modeled as the occultation of a circumstellar halo \citep{w06}. Another possible explanation is that the occulting disk edge is not infinitesimally thin but has some extension (an ``atmosphere") that results in some decline of the stellar brightness prior to the point of first contact at $\Delta$X$=-1$. This may be called the {\it fuzzy edge} hypothesis and a version of it has been modeled by \citet{sa08}. We propose an alternate explanation for the observed phenomenon below.

The lower left panel in Figure~\ref{fig4} shows that there is no significant change of V-I color of the system that depends on $\Delta$X in the range $\Delta$X$<-1.5$. However, as the lower right panel illustrates there is a  weak tendency for color to depend on brightness, a point first noted by \citet{Ham05}. This is expected because the real variations of star A, a weakly accreting T Tauri star, are caused primarily by the rotation of a spotted surface. Cooler spots are redder, of course, so we expect the star to be redder when fainter, as observed.  A linear least squares fit to the data yields a slope of 0.27 $\pm$ 0.10, which is representative of weak-lined T Tauri stars \citep{Her94}. 

Averaging the 71 measurements available when $\Delta$X$\leq-5$ we find a mean I magnitude for the out-of-eclipse state of this system of 14.456 $\pm$ 0.007. The star undergoes real variability (i.e. well in excess of the measurement error of a single point, $\sim$0.02 mag) that is normally within a range of 0.13 mag from I = 14.41 to I = 14.54, although on one night it was measured to be as faint as I = 14.58. The standard deviation of these 71 out-of-eclipse measurements is 0.06 mag. As noted above, there is no detectable color change with $\Delta$X in our data during the bright phases. The mean color based on the 87 measurements with $\Delta$X$\leq-1.5$ is V-I=1.575$\pm$0.004. 

For $\Delta$X between -5 and -1.5, we find a steady decrease in the brightness of the system characterized by a slope of 0.020 $\pm$ 0.003 mag R$^{-1}_A$. At the point of first contact, the system is  about 7\% fainter than when the star was more than 5 stellar radii above the disk ege. We show below (see Section 4.1) that there is a simple interpretation of this fact that does not involve either occultation of a stellar halo or invocation of a fuzzy edge. Rather, the observed decline in system brightness plausibly arises from a reduction in the contribution to the system light from a scattered component that depends on binary phase, not on $\Delta$X. 

\subsection{The Faint State of KH 15D}

Perhaps the biggest surprise of the recent monitoring has been the fact that the star has not faded to anywhere near the low brightness levels predicted by extant models. Also, as the 2009-2010 data clearly show, we continue to see substantial variation on the orbital period with a peak brightness of about I = 16.70. As Figs. \ref{fig3} and \ref{fig2a} show, this variation in the scattered light component has apparently existed as a ``floor" to the system brightness marked by the inflection point at $\Delta$X=+1 in the light curve for many years (see Section 3.1). If anything, the ``floor" was at a brighter level in the past (i.e. pre-2005) than it is today. The behavior of this component is not consistent with either a scattering halo around the individual stars or forward scattering (at least the simplest versions of such models), since they would predict a dependence of the scattered light on $\Delta$X, not on orbital phase. The rising brightness of the inflection point at $\Delta$X=+1 of the light curves seen on Figs. \ref{fig3} and \ref{fig2a} shows that the scattered light component is not a function of $\Delta$X.

Indeed, what the recent data show is that after all direct starlight is removed from the system, there is a persistent large-amplitude periodic variation that remains. In Fig. \ref{fig3a} we show the phased data from 2009-2010, which illustrates that variation, and a sixth-order polynomial fit. (We do not bother to fit the small, but real, reversal in brightness near phase 0.5 which is due to star B, since it is of no consequence to this discussion.) This rather substantial contribution to the system light has clearly been present since 1995 and may have been even brighter in the past. It is sufficiently bright that it contributes at about the 14\% level near a binary phase of 1.0 even when star A is fully visible.
  
\section{Discussion}

\subsection{Scattered Light}

It is clear that once star A is fully obscured we see a good deal of scattered light \citep{agol04,Ham05,w06,Her08,sa08} but the geometry of the scattering has remained uncertain. There are at least four possible scenarios, and they are not mutually exclusive. There may be a scattering halo around each star arising from matter within the magnetosphere, perhaps confined to accretion columns \citep{w06}. There may be forward scattering of light from each star reflecting off the presumed fuzzy edge of the occulting disk (or ring) \citep{sa08}. There must be some back scattered light from star A coming to us from the far side of the disk to account for the reflection effect observed in its velocity during eclipse \citep{Her08}. And, there could be general scattering from more widely dispersed dust in the vicinity but not closely associated with either star or the disk. In addition, it is possible that the optical depth of the occulting screen below its edge drops enough so that highly attenuated {\it direct} light from either or both stars is seen. And, it should be kept in mind that there may be a third (or more?) luminous object, most likely a low mass star, contributing to the system light. Here (see especially Section 4.4) we examine what the new data have to say about these possibilities.

A key fact, we believe, is that excess light depends on orbital phase and not on the location of star A with respect to the occulting edge, indicating that it cannot arise solely from either partial occultation of a scattering halo centered on star A or forward scattering from the fuzzy edge of the occulting disk. It could, however, arise from back scattering off the far side of the disk. We would naturally expect any such light to be modulated on the orbital period as the stars would illuminate the far disk differently depending on where they were in their orbits. While the total brightness of the scattered light is surprisingly high -- amounting to as much as 14\% of the light of the star -- this is perhaps not impossible, although it does require that the disk intercept a good fraction of the starlight radiated, even if the particles are nearly perfect reflectors.  

Regardless of the physical nature of this light, we can proceed empirically by simply assuming that a persistent periodic component on the orbital cycle, as revealed in the 2009-2010 data, has been present all along and remove it from the data when the system was bright to obtain a clearer picture of the occultation. The effect of doing this is shown in Fig. \ref{fig6}. In the left panel we show the pre-2005 light curve as a function of $\Delta$X, while on the right we show the brightness corrected for the periodic component based on the fit to it shown in Fig. \ref{fig3a}. The lines show the expected light curve for a ``perfect" knife edge -- perfectly sharp and perfectly opaque -- crossing a limb darkened star\footnote{The limb darkening prescription of \citet{sa08} was adopted here.} It is clear that such a model does not fit the measured data but does a reasonable job of fitting the corrected data. We infer that there is no longer any necessity to introduce scattering halos around the stars or a fuzzy edge to the disk to model the photometry. However, it is still probably necessary to have {\it some} direct or forward scattered light present in the system to explain all of the observations. In particular, the spectra analyzed by \citet{Her08} continued to show an absorption line component at the radial velocity of star A even after full occultation of the star. It persisted to at least $\Delta$X$\approx$4.  

\subsection{Spottedness of Star A}

The recognition of and correction for a periodic scattering component to the system light of KH 15D allows us to examine another aspect of its nature in more detail - its spottedness. While weak-lined T Tauri stars are known to be heavily and asymmetrically spotted from their periodic light variations, the possibility of detecting such features through occultation mapping is rare and, perhaps, unprecedented. Star A is known to be spotted and its rotation period of 9.6 days was determined by monitoring its light variations out of eclipse \citep{Ham05}. If the occulting edge is sharp enough, it might be possible to detect some asymmetry in its surface brightness related to its spottedness. 

Two features of the light and color curves of the star suggest that, in fact, there may be a broad asymmetry in the light distribution perpendicular to the direction of motion of the star relative to the occulting screen. It was recognized by  \citet{Ham05} that the system changes color quite rapidly as the occultation progresses -- much faster than one would expect if it were merely the bluish color of the reflected component that was to blame. If the leading edge of the occulting screen were to cover a more heavily spotted region of Star A first, this would lead to a general bluing of the system consistent with what is observed (see Fig.~\ref{Spot}).

The new data and interpretation reported here lend some support to such a view in the following way. It may be seen on Fig. \ref{fig6} that after correction for the scattered light component, the system's decline rate is actually somewhat delayed and then {\it steeper} than predicted by a completely sharp and opaque knife edge. A simple explanation for this is, once again, that the occulting edge encounters first a more highly spotted (and therefore less luminous) portion of the star and later passes across the hotter, brighter portion. The right panel of Fig. \ref{fig6} shows a very simple model of this, namely, a two component star in which 1/3 of the flux is radiated from the hemisphere first encountered by the occulting edge and 2/3 from the other. This flux ratio and implied temperature difference is consistent with what is known of the spots on weak-lined T Tauri stars \citep{Her94}. 

It is important to emphasize that we do not claim on the basis of Fig. \ref{fig6} to have actually detected the spottedness. We merely point out that the existence of this expected phenomenon, in concert with the periodic scattering correction, leads to a model light curve that is as good a fit to the data as one could expect. Note that for this discussion and in creating Fig. \ref{fig6} we have limited ourselves to the same set of pre-2005 data used to produce the models of \citet{w06} and \citet{sa08}. These models do not accurately predict the value of $\Delta$X for the more recent data, so we cannot directly include them on Fig.~\ref{fig6}. Also, there are clearly some variations from cycle to cycle in $\Delta$X due to variations in the location of the occulting edge, as discussed above. This leads to scatter in the measured data around the mean knife edge model and its effect is shown on the figures by the bracketing lines at $\Delta$X$=\pm0.5$~R$_A$. We turn now to a closer examination of the issue of variability in the precise location of the occulting edge on both short and long time scales.

\subsection{Smoothness of the Occulting Edge}

In the 2008-9 season it is clearly seen that during the third observed peak, the system was brighter than it was during the second and fourth and the maximum lasted longer (see middle right panel of Fig.~\ref{fig}). This indicates that the occulting edge did not extend as high up across the star as it had on the previous cycle and as it did on the next cycle. Clearly, the occulting edge does not advance uniformly or, equivalently, it has some structure -- bumps and wiggles, perhaps corrugation -- to it. To investigate this phenomenon, we have used the photometry during mid-ingress and mid-egress to pinpoint the location of the edge. At these times the brightness of the system is highly sensitive to the location of the edge with respect to the center of star A.

In Fig.~\ref{fig9} we show the difference between the computed value of $\Delta$X, based on the \citet{w06} Model 3 and the ``observed" value based on the knife edge model (corrected for the scattered light component) versus time. We chose Model 3 because it is most consistent with all of the observational and astrophysical constraints on the system. A positive deviation from zero on this plot indicates that the observed occulting edge was lower than expected for that cycle based on the model, i.e. that the system was observed to be brighter than the model predicted at that time. 

It is clear from this figure that the \citet{w06} Model 3 needs some revision to match the last five years of photometry. It has been predicting a more rapid advance of the occulting edge than actually observed. Most likely, a small adjustment to the orbits or to the model of the advancement of the edge will correct this, although a fuller revision is necessary to properly account for the scattered light component. Here we focus only on the scatter about the trend, which measures the degree of smoothness of the occulting edge on a timescale of the orbital period to a few years. These small scale fluctuations are not dependent on the details of the model used to compare with the observations.

The example of 2008-9 appears as the last set of data points on this figure around JD=2454800. It is evident that the third peak of this set is about 0.4 stellar radii lower than the second or fourth peaks, quantifying the phenomenon noted above. This degree of variation within a single observing season, normally comprising four orbital cycles, is clearly characteristic of the star, as the eight years of intensive photometry displayed reveal. The standard deviation of the data around the mean trend is about 0.25 stellar radii. Some of this is due to photometric errors, of course, but the stellar brightness drops so rapidly in the regime to which we have limited the exploration ($-0.3<\Delta$X$<0.6$) that the expected scatter from photometric variations characteristic of the star out of eclipse, is 0.15 stellar radii or less. This suggests that the true cycle-to-cycle variations of the occulting edge are characterized by a standard deviation of about 0.2 stellar radii, or a mere 1/10 of a stellar diameter!

One may ask whether this small, but measurable, variation in the location of the occulting edge from cycle to cycle is fully responsible for the scatter about the mean knife-edge model or whether other physical phenomena are required. To address that question, we show, in Fig.~\ref{fig13} all of the CCD data obtained during an egress or ingress. For the post-2005 data, the plotted value of $\Delta$X was obtained by correcting the predicted value from \citet{w06} for each ingress/egress pair by a small amount based on the analysis of Fig.~\ref{fig9}. For the pre-2005 data a single correction value of -0.1 was applied to all of the data. On each panel, two models are shown, one being a standard limb-darkened star occulted by an opaque, sharp knife edge (dashed line). The other is identical except that the hemisphere first occulted radiates 1/3 of the total luminosity of the star while the other radiates 2/3. This is referred to as the spotted case (see Fig.~\ref{Spot}). It is clear that when allowance is made for the scattered light component, a slightly variable height in the occulting edge, and perhaps a spotted surface, the data fit the knife edge model very well.

\subsection{The Source of the Scattered Light}

As described above, the recent photometry indicates that there has likely been, all along, a periodic (on the orbital cycle) and surprisingly bright underlying component contributing to the brightness of the KH 15D system (Fig. \ref{fig3a}). Here we consider the possibility that it arises at least in part from back scattered light off the far side of the same circumbinary disk that is responsible for the occultation of the orbits. Recall that such back scattering is required by the reflection effect observed in the radial velocity of star A \citep{Her08}.

At maximum brightness, the reflected component has I=16.7, which is 2.1 mag fainter than Star A's inferred brightness of I=14.6. If all of the scattered light comes from this single source, then the disk would need to intercept about 14\% of the solid angle into which the star is radiating assuming perfect reflectors to produce something this bright. Clearly such a reflecting surface would need to be of high opacity, extended and significantly warped. These, of course, are the same conditions that must exist to account for the occulting function of the same disk and its slow variation with time, attributed to precession \citep{CM04}.

It is also possible that this reflected component has been brighter in the past than it is today. It may be seen on Figs.~\ref{fig1} and \ref{fig3} that during minimum light, the system was apparently brighter in the late 1990's and early 2000's than it is today. This is not unexpected qualitatively since the same gradual precession that is causing the occulting edge to advance would be changing the solid angle of the disk as seen from Earth. Some slow evolution of the scattered component would be expected in this hypothesis. While a detailed model of such a scattering/occulting disk is beyond the scope of this paper its exploration does seem warranted by the observations.

Alternatively, or in addition, one should consider the possibility that the ``opaque" edge of the occulting material is currently somewhat sharper than it was in the past. As noted previously, spectra taken about a decade ago indicate that there is some direct light seen even when star A is several radii below the occulting edge. And, the general system brightness was higher at that time, probably at all binary phases. For the past five years, however, a perfectly sharp and opaque occulting edge provides an excellent model for the photometry, as shown in Fig.~\ref{fig13}. Perhaps as the disk has precessed it has become increasingly more optically thick owing to geometrical effects. This could be tested with a modern epoch high resolution spectrum that would show little or no evidence of light at the velocity of star A at any phase if the disk is opaque and forward scattering minimal or absent.

\subsection{Color Variations}

A final interesting and unexpected result from the current data set is the clear evidence for color variations during minimum light. What factors could be involved in producing these? Two possibilities exist -- either one or more sources of the light in KH 15D is highly variable or the occulting/scattering geometry alone produces this behavior. We consider first the possibility of a variable source. Star A shows only small light and color variations out of eclipse, so if it is involved then we would need to invoke a binary phase dependent variation, such that the star became much more active near periastron than near apastron. This is not unprecedented; other binary T Tauri stars, such as DQ Tau, have been thought to suffer {\it pulsed accretion} events associated with perihelion passage \citep{Bas97}. If something like this is happening in KH 15D it would possibly affect Star B as well, since it is presumably very similar in mass and luminosity to star A and is believed to be the dominant source of the scattered light during the central minima. 

In support of a picture such as this, one can see that in the 2007-8 season (middle left panel of Fig. \ref{fig}) the third minimum was very blue and this also corresponded to a relatively high and sharp central reversal in the light curve at a time when Star B was the primary light source. If there were an outburst of the star, perhaps associated with pulsed accretion near periastron, then the brightness and blueness of the star may well account for the changes observed. In the following cycle, one sees that the light curve is not nearly so peaked and bright, although the central reversal is still obviously present, and the star becomes redder during this phase.

While variations of star A or B might explain excursions to the blue, it does not seem likely that they can, by themselves, explain the red events. Flares on T Tauri stars, whether generated by surface magnetic field reconnections or by accretion events, involve a brightening associated with a heating of the photosphere that causes the star to get bluer. It would be surprising for star A or B to become intrinsically redder for any reason to the degree required (0.4-0.5 mag). It would also be very unexpected for star B to be intrinsically redder than star A, since it is known to be brighter and, assuming coevality, should be more massive and, therefore, bluer. One explanation might be that star B's light is, at times, highly reddened by dust. If so, this would constitute the first clear evidence for small particles within the KH 15D system.

If the system redness is reflecting an actual drop in the color temperature of an illuminating source, it seems more likely that this would be coming from a third source of light -- a possible faint, red companion. The existence of a third light within the system, from either a faint star, brown dwarf or perhaps even young giant planet, could also account for the tilt and warping of the disk relative to the binary plane that is required to explain the observed properties of the KH 15D circumbinary disk.  Perhaps it is the redness of this putative object that leads to the color variation of the system at times when star A and B are not contributing enough either by direct or scattered light. As noted previously, \citet{p08,p10} suggest that KH 15D may be a member of a class of variables whose properties depend on their being three-body systems.
     
The other class of possibilities is that the occulting/scattering geometry may be constantly changing as the two known luminous objects (stars A and B) within the system orbit. It is conceivable that changes in that geometry alone could be responsible for the observed color changes. Reddening can occur if small grains are present along some lines of sight within the system and/or from reflection off of rocky or weathered surfaces. Bluing of the light can likewise arise from viewing the scattered component produced by small grains or by reflection off some solids, such as fresh water ice. The variations may be caused by variable illumination of zones within the disk that have different scattering properties. For example, the blueness may result from reflection off of fresh ice, while the redness may come from reflection off of silicates or dirty ice. Clearly one will require further investigation of this phenomenon, including infrared spectral studies to resolve this new mystery of the KH 15D system.

\section{Final Remarks and Summary}

Five additional years of photometric data have been added to the growing record on KH 15D. Most of this is obtained in the I band but we have also now begun regular monitoring in V as well. This paper has presented, examined and discussed these data. We have also provided a basic interpretation of the wealth of information encoded in the photometry, with emphasis on the newly discovered aspects of its behavior.

The data presented here indicate that likely, all along, there has been an underlying component of presumably scattered light with a large amplitude periodic variation on the orbital cycle present in the KH 15D photometry. It was not noticed previously because of the more dramatic eclipse behavior and its relative faintness compared to direct light from Star A. Correcting for this component indicates that the apparent decline in system brightness that was thought to precede the time of first contact is much weaker or absent in the data (see Fig.~\ref{fig6}). This in turn means that there is no urgent requirement for a scattering halo around the individual stars or a fuzzy edge to the disk which results in forward scattering to model the data. In addition, neither of these scattering models is consistent with the observed underlying component, in particular its dependence on orbital phase rather than on $\Delta$X. Some direct light or forward scattering is necessary (or was necessary) in the system to explain the radial velocity components described by \citet{Her08} in spectra obtained when star A was fully occulted.

The other major surprise from the last five years of monitoring the KH 15D system is the extreme color changes that are now well documented for the star during times of central minima. Sometimes the system becomes bluer during these phases and sometimes redder, with swings of 0.4 mag in V-I color common. This discovery raises the possibility that we are detecting the light of a third luminous object (faint star) which is very red or that there is an inhomogeneous set of scattering objects. 

A working hypothesis is that all of the photometry and spectroscopy obtained so far can be understood in terms of the following scenario. The binary orbits within a tilted, warped and precessing disk or ring that stretches from 0.6-5 AU or so and is extremely thin. It is composed of mm or cm-sized grains of presumably silicate composition close to the stars and ices further out that have settled into a very thin, essentially gas-free plane \citep{l10}. Some gas is still accreting through the disk since the system continues to drive a jet/outflow \citep{M10}. The disk is optically thick perpendicular to its plane and must be warped in such a way that its front edge now blocks the orbits of the stars while its back side manages to reflect a good deal of light in our direction. This light produces the reflection effect in the radial velocity of star A \citep{Her08} as well as the periodic scattering component described in this paper.

If this picture is basically correct, it suggests that a phase in the evolution of planetary systems very much like the one envisioned originally by \citet{GW73} exists. Namely, a dynamically independent ``sub-layer" of solids has formed. If it exists in KH 15D, then it is likely to be a common feature of T Tauri stars and a relatively stable phase of planetary evolution. If this disk follows the evolution outlined by  \citet{GW73} and more recently by \citet{CY10} we may expect gravitational fragmentation into asteroid-sized object in the not-too-distant future.

Confirmation of this general picture will require additional work on this system during the current phase when both stars are obscured. Clearly it will be important to continue the optical photometry and to ascertain the causes of the large color variations. In addition, it would be very useful to add near infrared monitoring and both low and high dispersion infrared spectroscopy to search for a third light component and to study the nature of the reflecting objects. Polarization measurements would surely be helpful in constraining the nature and geometry of the scattering objects. Finally, a new set of models based on our improved understanding of the scattered light and the (now) complete set of data on the occulting phase appear warranted.  

\acknowledgments

We thank the large number of Wesleyan undergraduate and graduate students who participated in obtaining the observations at VVO reported here. This work was partially supported by a NASA grant through its Origins of Solar Systems program to one of us (W.H.)

\clearpage

\begin{deluxetable}{cccc}
\tabletypesize{\scriptsize} \tablecaption{Photometry Parameters for the 2005-9 Seasons. \label{PhotParm}} \tablewidth{0pt} \tablehead { \colhead{Observatory}   & \colhead{Aperture Radius} & \colhead{Inner Annulus} & \colhead{Width}}

\startdata

VVO 05/06 & 6.5 (3.9\arcsec) & 10.5 (6.3\arcsec) & 2 (1.2\arcsec)\\
VVO 05/06 & 11(3.7\arcsec) & 15 (5.1\arcsec) & 2 (0.7\arcsec)\\
KPNO & 4.2 (2.5\arcsec) & 10.5 (6.3\arcsec) & 2 (1.2\arcsec)\\
CTIO & 5-6 (1.8\arcsec - 2.2\arcsec) & 13 (4.8\arcsec) & 2 (0.7\arcsec)\\
MMO (1.5 m) & 10 (2.7\arcsec) & 11 (3.0\arcsec) & 5 (1.4\arcsec)\\
MMO (0.6 m) & 7 (4.7\arcsec) & 8 (5.4\arcsec) & 5 (3.4\arcsec)\\

\enddata
\end{deluxetable}

\begin{deluxetable}{ccc}
\tabletypesize{\scriptsize} \tablecaption{Adopted V and I magnitudes for Comparison Stars \label{CompMags}} \tablewidth{0pt} \tablehead { \colhead{Star}   & \colhead{V} & \colhead{I}}

\startdata

A & 13.37 & 12.55 \\
C & 12.97 & 12.24 \\
E & 13.885 & 12.61 \\
F & 13.87 & 12.90 \\
16D & & 14.30 \\
17D & & 15.16 \\
31D & & 14.41 \\

\enddata
\end{deluxetable}

\begin{deluxetable}{ccccccc}
\tabletypesize{\scriptsize} \tablecaption{Photometry of KH 15D from 1995-2009\label{Data}}  \tablewidth{0pt} \tablehead { \colhead{JD}   & \colhead{I} & \colhead{$\sigma$} & \colhead{V-I} & \colhead{$\sigma$} & \colhead{$\Delta$X} & \colhead{Obs.}}

\startdata

2450017.8375 & 14.449 & 0.011 &  9.999 & 9.999 & -5.79 & VVO \\
2450021.7443 & 14.521 & 0.024 &  9.999 & 9.999 &  -2.69 & VVO \\
2450026.7416 & 17.132 & 0.074 &  9.999 & 9.999 &   2.60 & VVO \\
2450028.7966 & 14.562 & 0.010 &  9.999 & 9.999 &   4.86 & VVO \\ 
2450031.7696 & 14.014 & 0.007 &  9.999 & 9.999 &   6.10 & VVO \\
\enddata
\tablecomments{Times are geocentric, not baryocentric. Missing data is indicated by the value 9.999. Table 3 is available in its entirety in machine readable form.} 
\end{deluxetable}

\begin{figure}
\epsscale{1.0}
\plotone{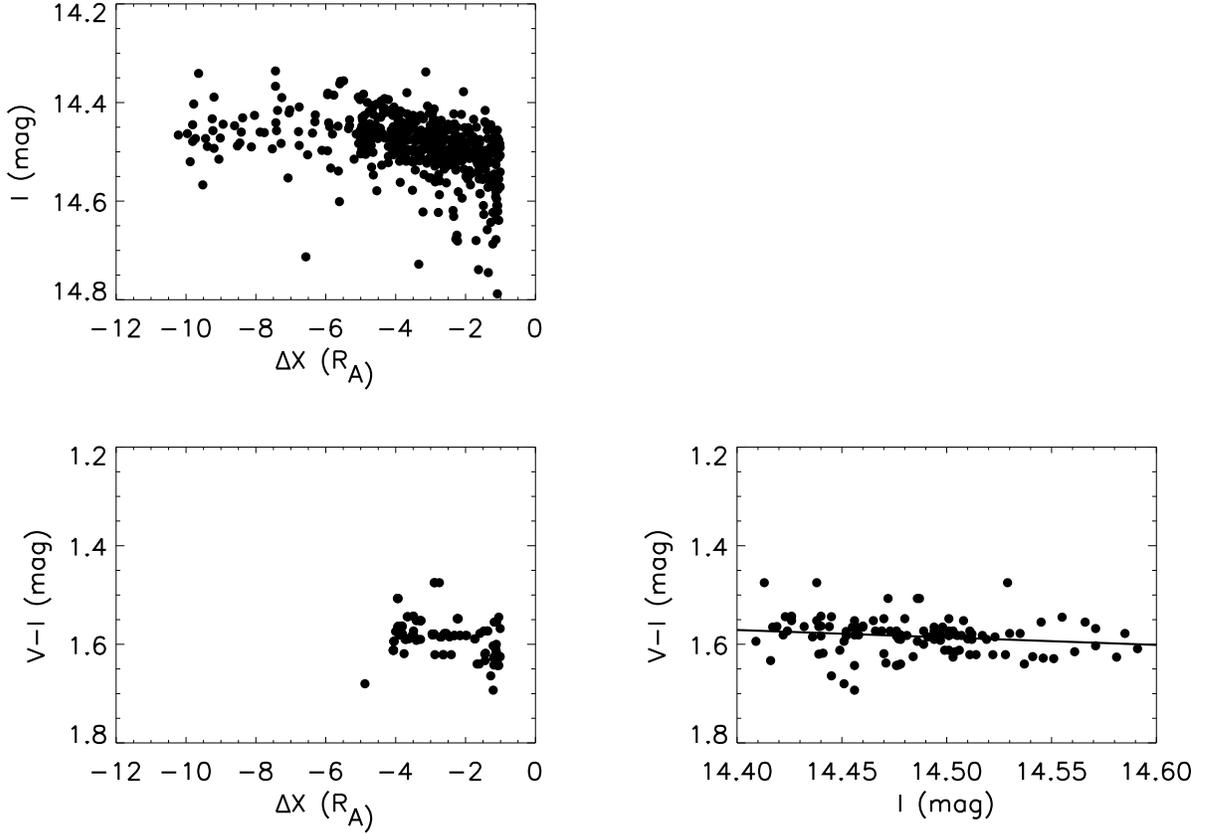}
\caption{The top panel shows that when Star A is more than 5 stellar radii above the disk edge it shows no dependence of brightness on $\Delta$X. However, for $-4.7 \le \Delta{\rm X} \le -1$ there is a clear dependence of brightness on height above the disk, characterized by a slope of 0.020 $\pm$ 0.003 mag per stellar radius. The lower left panel shows that there is no detectable reddening of the system associated with the observed decline in brightness for $-4.7 \le \Delta{\rm X} \le -1$. The lower right panel shows that when the star is in its non-eclipsed state ($\Delta$X $<$ -1.5, there is a small but significant correlation between its brightness and color, as reported previously by \citet{Ham05}. This is due to the fact that the brightness variations out of eclipse are caused by the rotation of a spotted star.}
\label{fig4}
\end{figure}

\begin{figure}
\epsscale{1.0}
\plotone{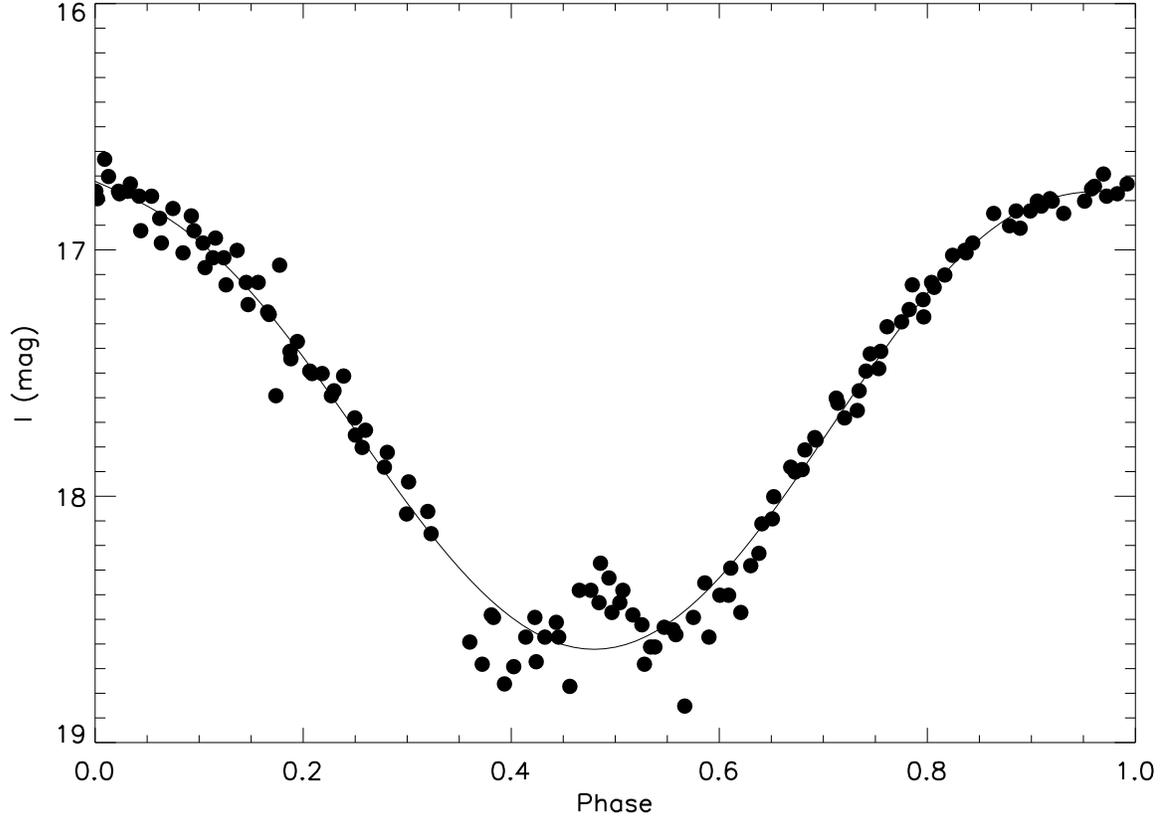}
\caption{The 2009-10 data are plotted (solid circles), phased with the orbital period of 48.37 days. A sixth degree polynomial fit is shown (solid line) and was used to correct the total system light for the scattered light component. Note that the reversal of the decline near phase 0.5 is real and is caused by the rise of star B to a point where it is the dominant source of the reflected light received at Earth during those phases. This small reversal was not included in the fit to the data since it has no effect on the analysis done in this paper.}
\label{fig3a}
\end{figure}

\begin{figure}
\epsscale{1.0}
\plotone{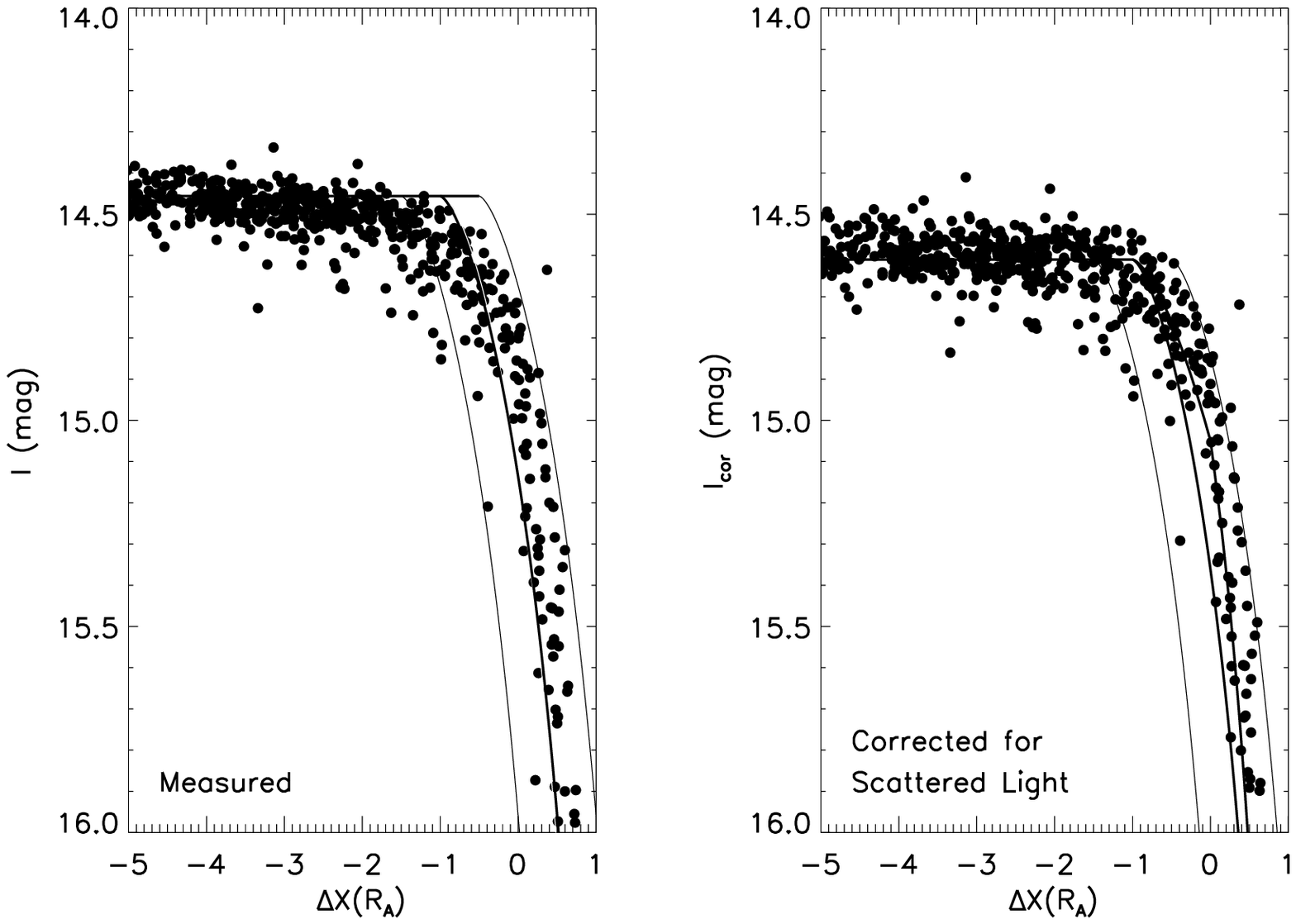}
\caption{The left panel shows the pre-2005 data \citep{Ham05} plotted versus $\Delta$X, the distance of the center of star A from the occulting edge in units of the stellar radius based on the model of \citet{w06}. It is clear, as noted by \citet{w06} and \citet{sa08}, that there is an apparent decline in brightness of the system that occurs well before the point of ``first contact" at $\Delta$X=-1. When account is taken of the scattered light (by subtracting a fit to the 2009-10 light curve from all earlier data, as shown in Fig. \ref{fig3a}), however, one obtains a light curve that has no such precursor (right panel). The solid lines on both panels show the expected light curve for a limb darkened star occulted by a knife edge of negligible thickness with no forward scattered light or halo. The limb darkening prescription of \citep{sa08} has been adopted. Three lines are shown corresponding to $\Delta$X=(-0.5,0,+0.5) allowing for some range in this quantity. A fourth line on the right hand panel (third line from left) represents the spot model described in the text. }
\label{fig6}
\end{figure}

\begin{figure}
\epsscale{0.5}
\plotone{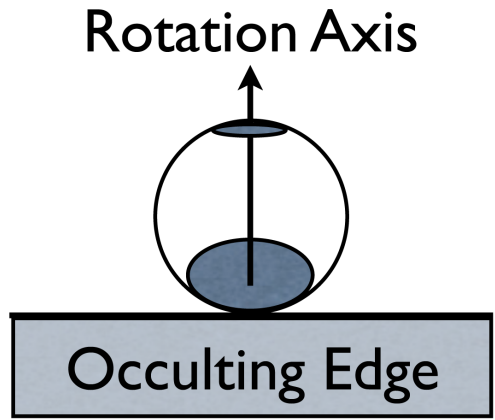}
\caption{Schematic diagram showing the orientation of the spots on KH 15D and the occulting edge that is inferred from our data for the time of first contact. The star ``sets" behind the occulting edge by sinking vertically on this schematic diagram. Since the cooler part of the star (the spot) is covered first, the system remains, at the beginning of the eclipse, a bit brighter and bluer than it would otherwise have been, and then it drops in brightness more rapidly and changes color in a manner unlike a star without spots.}
\label{Spot}
\end{figure}

\begin{figure}
\epsscale{1.0}
\plotone{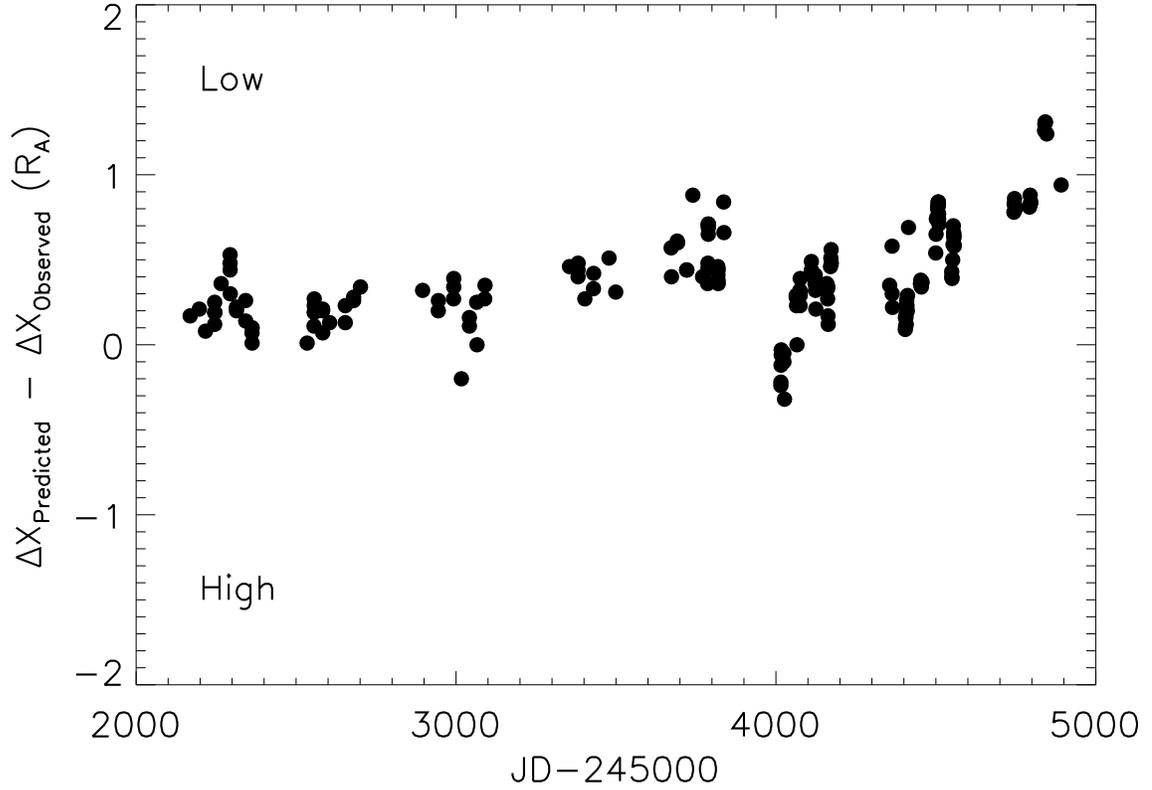}
\caption{The difference between $\Delta$X predicted from Model 3 of \citet{w06} and ``observed" (based on the knife edge model corrected for scattered light) is plotted against Julian Date. A positive value of this quantity indicates that the system was brighter than Model 3 predicted at that time, meaning that the occulting edge was observed to be lower with respect to the star (i.e. the system was brighter) than the prediction. The unit on the y-axis is the radius of star A (1.3 R$_\odot$). Only data for which the value of $\Delta$X inferred from the photometry lay within the range $-0.3<\Delta$X$<0.6$ are plotted, because these are the most reliable points, corresponding to the times when the system brightness is changing most rapidly.}
\label{fig9}
\end{figure}

\begin{figure}
\epsscale{1.0}
\plotone{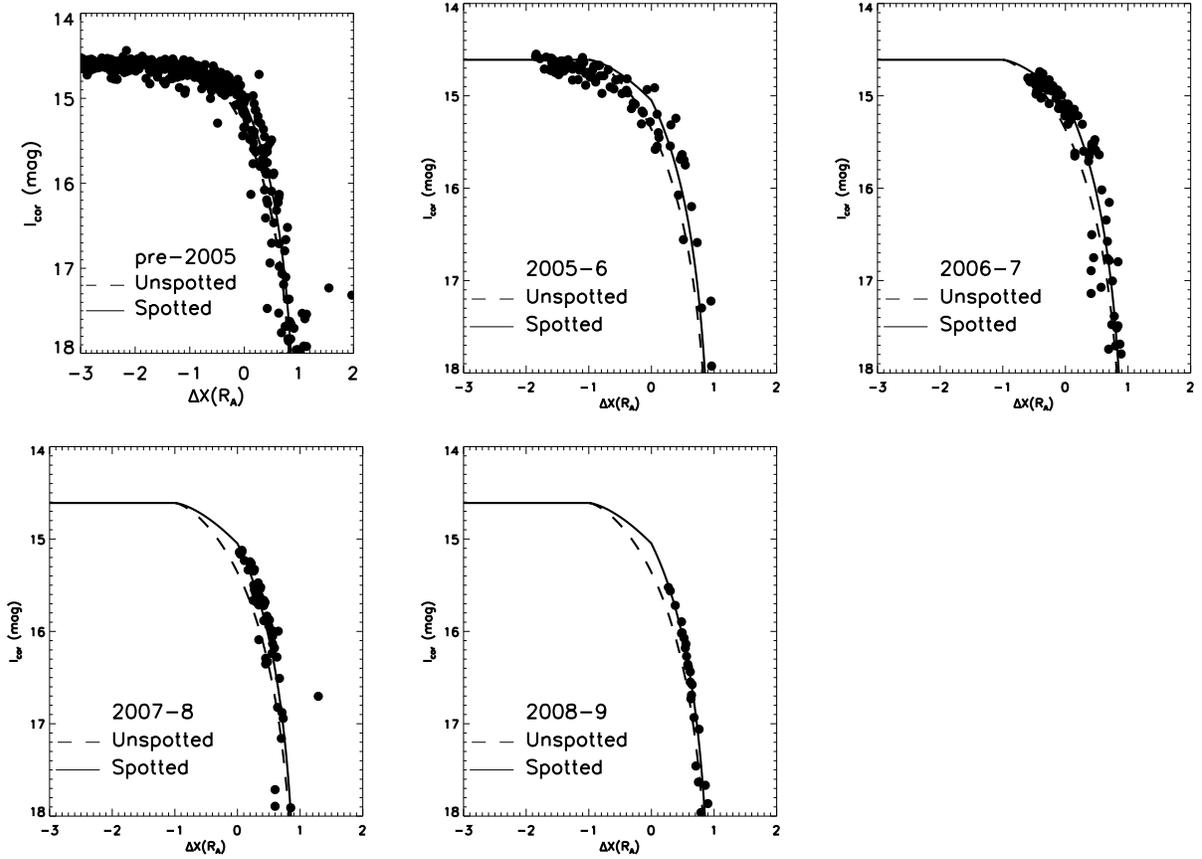}
\caption{The pre-2005 data from \citet{Ham05} and recent data, by observing season, from this paper are shown corrected for scattering as described in the text. It is clear that when allowance is made for the scattered light component, a slightly variable height in the occulting edge, and perhaps a spotted surface, the data fit the knife edge model very well.}
\label{fig13}
\end{figure}

\end{document}